# Haptic in-sensor computing device made of carbon nanotube-polydimethylsiloxane nanocomposites

Kouki Kimizuka[1], Saman Azhari[1,2,3,*], Shoshi Tokuno[1], Ahmet Karacali[1], Yuki Usami[1,3], Shuhei Ikemoto[1,3], Hakaru Tamukoh[1,3], Hirofumi Tanaka[1,3,*]

*Corresponding author: saman@aoni.waseda.jp, tanaka@brain.kyutech.ac.jp

[1]Graduate School of Life Science and Systems Engineering, Kyushu Institute of Technology, 2-4 Hibikino, Wakamatsu, Kitakyushu 808-0196, Japan

[2]Graduate School of Information, Production and Systems (IPS), Waseda University, 2-7 Hibikino, Wakamatsu, Kitakyushu, Fukuoka 808-0135, Japan

[3]Research Center for Neuromorphic AI Hardware, Kyushu Institute of Technology, 2-4 Hibikino, Wakamatsu, Kitakyushu 808-0196, Japan

Abstract

The importance of haptic in-sensor computing devices has been increasing. In this study, we successfully fabricated a haptic sensor with a hierarchical structure via the sacrificial template method, using carbon nanotubes-polydimethylsiloxane (CNTs-PDMS) nanocomposites for in-sensor computing applications. The CNTs-PDMS nanocomposite sensors, with different sensitivities, were obtained by varying the amount of CNTs. We transformed the input stimuli into higher-dimensional information, enabling a new path for the CNTs-PDMS nanocomposite application, which was implemented on a robotic hand as an in-sensor computing device by applying a reservoir computing paradigm. The nonlinear output data obtained from the sensors were trained using linear regression and used to classify nine different objects used in everyday life with an object recognition accuracy of >80 % for each object. This approach could enable tactile sensation in robots while reducing the computational cost.

## Introduction

As artificial neural networks (ANN) continue to diversify due to the rapid expansion of artificial intelligence (AI), the knowledge that was once a niche interest has become essential. The development of in-sensor computing devices is a direct outcome of the growth of AI in various fields. These devices, with their unique capabilities, are poised to play a significant role in the future of AI, particularly in the management of big data and real-time computational scenarios[1–4]. Some of these devices have been explored for their potential to incorporate optoelectronic components, enabling the creation of microscale and energy-efficient neuromorphic devices using complementary metal-oxide-semiconductor technology.[5] The concept of in-sensor computing, which integrates sensing and computing functions, suggests that material-based sensors can perform computing tasks. This integration, unlike traditional setups where sensing and computation are distinct, reduces power consumption by minimizing the need to shuttle information between the processing and memory components, especially when dealing with large datasets. As a result, these devices hold promise for effectively managing big data, particularly in real-time computational scenarios. Our study's findings on the haptic in-sensor computing device contribute to this promising field, offering practical applications in the management of big data in real-time computational scenarios.

The concept of in-sensor computing is similar to that of in-materio reservoir computing (RC). RC is a computational framework derived from a recurrent neural network in which electrical signals are fed into a random network of nonlinear nodes, and the outputs are linearly combined and computed using linear regression[6]. The RC framework enables physical matter[7–10] through hardware implementation, leveraging its inherent principles of recurrent dynamics and reservoir computing architecture. This ability is attributed to the theory behind the RC, which allows the internal state of the reservoir (hidden layer) to remain unchanged while the output signal is trained using simple linear regression. This poses the question of what a reservoir could be. A network of porous carbon nanotubes (CNTs) embedded in polydimethylsiloxane (PDMS) serves as a pressure reservoir capable of performing classification and recognition tasks. Specifically, the CNT-PDMS sensor exhibits the potential to discern various stiffness levels, effectively functioning as an in-sensor computing device leveraging material properties. The feasibility of achieving efficient classification with high accuracy solely based on haptic time-dependent data prompts inquiry into pertinent parameters and their application to enhance sensor performance. Addressing these inquiries could position the sensor as a haptic in-sensor computing device applicable across diverse domains, including soft and sensitive pressure sensors for artificial skin applications, as well as seamless integration into wearable devices for medical, nursing, and welfare applications[11–18], particularly in sensory prostheses and soft robotics. [19–21] Addressing a gap in current knowledge, this study builds upon our prior work demonstrating the potential of CNT–PDMS nanocomposites in agricultural and robotics applications, particularly in classifying tomato maturity. We have previously showcased the fabrication of flexible and highly sensitive tactile sensors by harnessing the synergistic properties of highly conductive CNTs and flexible PDMS with a low Young's modulus. To further advance this research, we investigate the capacity to emulate human skin-like performance in a pressure sensor configuration by integrating soft materials, exemplified by CNT-PDMS nanocomposites, with coordinate electrode arrays. In this study, we aim to explore the potential of coordinate electrode arrays in enhancing the detection of pressure differences across various points, thereby improving the sensor's ability to discern subtle variations in tactile stimuli. The novelty of our approach lies in the integration of the fabricated nanocomposite with a matrixed-electrode array on a robotic hand, facilitating the collection of multiple outputs from distinct points on the sensor. These outputs are subsequently leveraged to perform object recognition using a novel RC model, thereby enabling accurate classification of the products grasped by the robotic hand (**Fig. 1**).

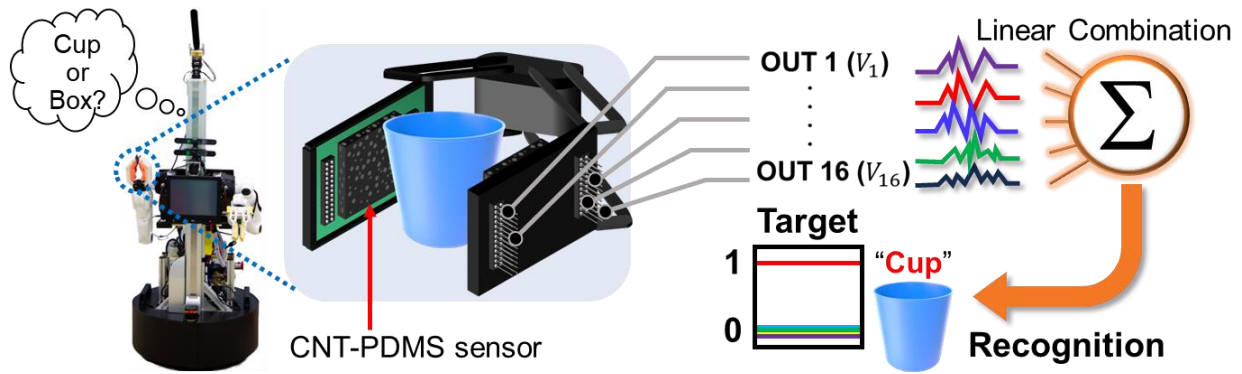

**Fig. 1** Schematic illustration of haptic in-sensor computing in which the robot predicts "cup". Haptic sensing data has been collected and applied to the reservoir computing paradigm by updating the weight using linear regression.

## Results

The scanning electron microscopy (SEM) image (**Fig. 2(a)**) of the fabricated CNT-PDMS sensor confirms the formation of a porous structure and the complete removal of the sugar particles. The piezoresistive properties of the samples with different CNT contents (0.2 g, 0.3 g, 0.4 g, and 0.5 g) were quantitatively evaluated using the setup shown in the method section. Each measurement was conducted five times to determine the mean and standard deviation of the piezoresistive performance of the devices. Subsequently, we evaluated the sensitivity and elastic modulus based on the obtained data. In piezoresistive pressure-sensing devices, sensitivity is an essential parameter that describes the change in the output signal relative to the initial output signal in a specific pressure range[22]. For a pressure sensor, the magnitude of the change in the output corresponding to a certain pressure range is important. Therefore, a large change in resistance indicates high sensitivity, whereas a small change in resistance indicates low sensitivity. First, we present the results of resistance versus force for up to 5 N, as illustrated in **Fig. S1**. As shown in **Fig. S1(a),** we confirmed the resistance changes induced following the application of a force based on the piezoresistive effect. The sample with 0.2 g of CNTs exhibits a decrease in resistance by 6.25 kΩ in total, as shown in **Fig. S2**. The resistance changed notably in the force range of 2–3 N. Moreover, the sample with a 0.3 g CNT (see **Fig. S1(b)**) shows a decrease in resistance when a force was applied (3–5 N). Despite the minimal change in the number of CNTs, a significant difference in sensitivity was observed. In contrast, **Figs. S1(c)** and **S1(d)** both display linear decreases in resistance compared with those observed in **Figs. S1(a)** and **S1(b)**. To understand why this variation occurs, we studied the sensor's mechanical behavior because the stiffness of the CNT-PDMS sponge-like sensor directly affects the sensitivity. **Fig. 2(b)** shows the elastic moduli of the samples with different CNT contents. The elastic behavior trend changed from nonlinear to linear when the CNT content increased. When a force was applied to the sensor, the sample with a low-CNT content exhibited a nonlinear curve owing to its high elasticity, which is a property of PDMS. In contrast, the sample with a high CNT content exhibited a linear trend owing to its rigidity, which resulted from the mechanical strength of the CNTs (**Fig. S3**). The change in resistance of the porous CNT-PDMS nanocomposite was attributed to the replacement of air-filled pores by resistive CNT-PDMS because of the applied pressure. This occurrence, particularly the samples with 0.2 g and 0.3 g, suggests an exponential tendency when force is continuously applied (**Figs. S3(a)** and **S3(b)**), resulting in nonlinear changes. **Fig. 2(c)** shows the sensitivity curves of the CNT-PDMS sensors at different CNT contents. Sensitivity $S$ is represented as the relative change in resistance $|\Delta R/R_0|$ with respect to the relative change in pressure $\Delta P/P_0$ (**Equation (1)**).

$$S = \frac{\partial(\Delta R/R_0)}{\partial(\Delta P/P_0)} \quad (1)$$

where $R_0$ and $P_0$, respectively, denote the initial resistance and pressure values[23]. These values were calculated and plotted using the mean data, as shown in **Fig. S1**. As a result, the sample synthesized with 0.2 g yielded the highest sensitivity ($S = 0.94$) when the relative change in pressure was in the range of 40–60 %. Furthermore, an increase in the CNT content reduced the sensitivity of the piezoresistive device owing to the formation of conductive paths in the porous structure and an increase in the rigidity of the nanocomposites. In the high-CNT content sensors (0.4 g and 0.5 g), the resistance changes were minimal because of the mechanical strength of the CNTs, which affected the stiffness of the sensor. Based on these results, samples with a lower CNT content exhibited nonlinear behavior, which is a crucial property of reservoir-computing devices. **Fig. 2(d)** shows the time series data of the resistance change in the fabricated CNT-PDMS nanocomposite with 0.2 g CNT content after repeated loading and unloading. Response time ($\Delta t_{res}$) and recovery time ($\Delta t_{rec}$) were calculated from the difference between the maximum and minimum resistance values at three locations for each measurement (**Fig. S4**). **Fig. 2(e)** shows the response and recovery time results for all measurement samples. The four different samples with CNT contents of 0.2 g, 0.3 g, 0.4 g, and 0.5 g yielded the values of $\Delta t_{res}$ of $0.35 \pm 0.068$ s, $0.41 \pm 0.031$ s, $0.51 \pm 0.17$ s, and $0.54 \pm 0.22$ s, while the values of $\Delta t_{rec}$ were $0.16 \pm 0.057$ s, $0.13 \pm 4.8 \times 10^{-4}$ s, $0.13 \pm 0.047$ s, and $0.13 \pm 4.3 \times 10^{-6}$ s, respectively. The resistance in samples with CNT contents of 0.4 g and 0.5 g took longer to reach full saturation owing to their rigidity, while the recovery time in all samples was less than 0.25 s. In short, samples with lower CNT content exhibited high sensitivity and elasticity, fast responses, and short recovery times, which implies that these samples have the potential for use in piezoresistive sensing devices and reservoir sensing performance. Finally, we obtained a sensor output (>10000 cycles), as shown in **Fig. 2(f)**. As cycling progressed, the resistance of the sample decreased until it reached a point of stabilization. This decrease in resistance can be attributed to the enhanced interconnections among the CNT facilitated by the shrinkage of PDMS during the process. There were no significant resistance changes throughout the measurement, indicating that the fabricated sensors generated a stable output and exhibited excellent stability and endurance.

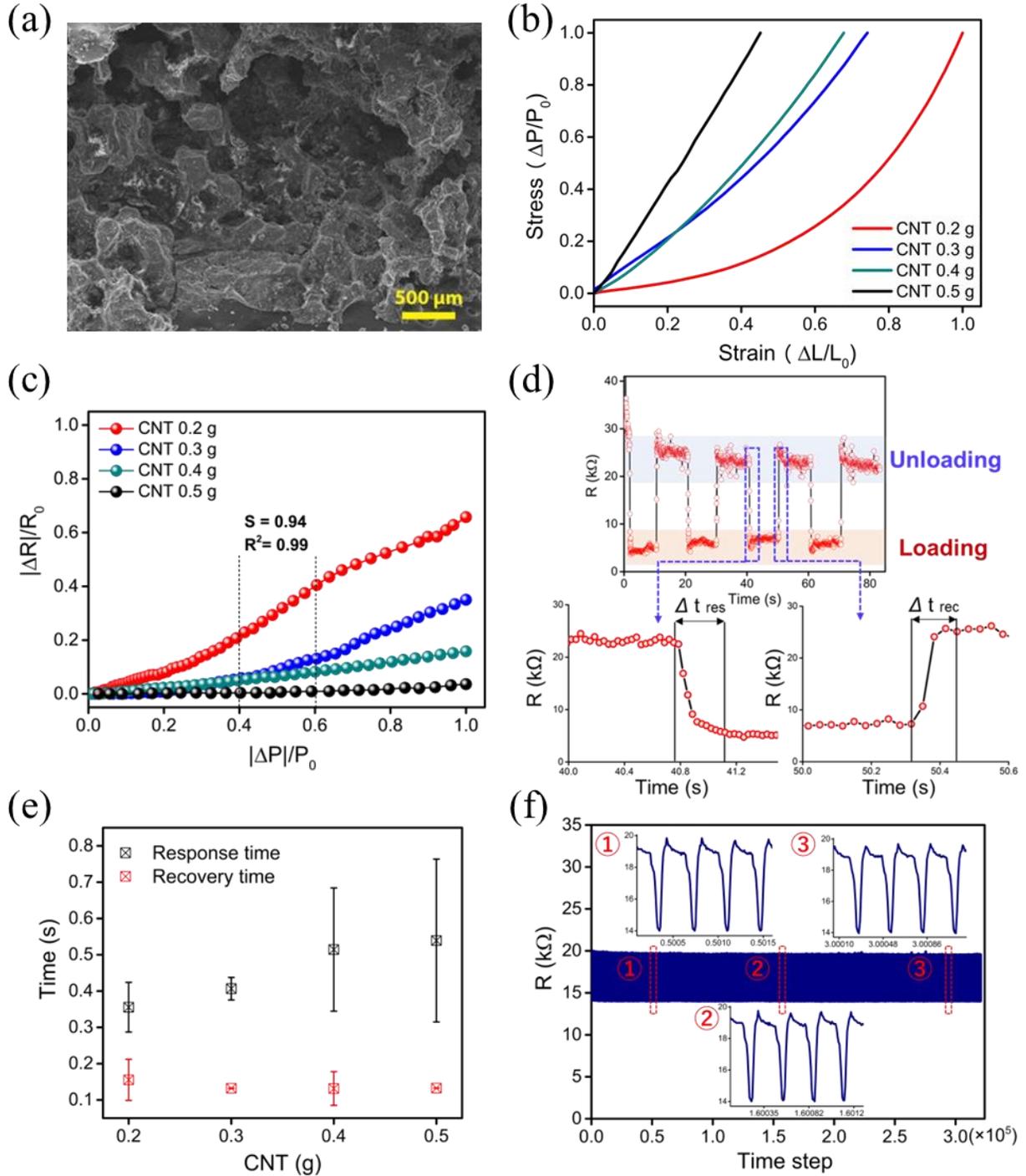

**Fig. 2** (a) Scanning electron microscopy (SEM) image of CNT-PDMS sensor (Scale: 500 μm). (b) Elastic modulus and (c) sensitivity of samples with different amounts of CNT. (d) Time series measurement of repeated loadings and unloadings in the sample with 0.2 g CNTs (e) Response and recovery time of CNT-PDMS nanocomposites (0.2–0.5 g). (f) Change in resistance during 10000 cycle tests with CNT-PDMS nanocomposites (0.3 g). The graph above represents four cycles at three different cycles.

We demonstrated object recognition using the CNT-PDMS nanocomposites sensor as a haptic "in-sensor" computing device. The in-sensor computing device was mainly developed as a vision device,[1,24], whereas haptic devices have not been extensively developed. To implement the fabricated sensor in the robotic hands, the sample was attached to the printed circuit board (PCB) via a silane coupling agent (**Fig. 3(a)**) to minimize the contact resistance between the sensor and the PCB board[25]. **Fig. 3(b)** shows a schematic of the object recognition setup, which was implemented by using a robotic hand for dynamic time-series data acquisition for the classification of nine objects (**Fig. S5**). For these measurements, we prepared a PCB with 49 electrodes in a matrix (**Fig. S6**) and used a sample with a CNT content of 0.3 g. Because the reservoir systems were affected in a time-dependent manner, the sample with 0.3 g CNT was selected for this task owing to its small deviation in response and recovery times. Based on the temporal dynamics, the sample with 0.3 g CNT was selected for this task because of its small deviation in the response and recovery times. For object recognition, we prepared various objects (such as boxes and plastic cases of different sizes) with different shapes and hardness values. Training and classification were conducted through a single hot-vector representation using linear regression for object classification using a robotic operating system (ROS). The output was divided into 50 % training data and 50 % testing data. For this measurement, the number of data points was set to 100, with a sampling rate of 10. We must note that classification is considered successful when the average likelihood of the target object is higher than that of the other objects. As shown in **Fig. 4(a)**, dynamic data were obtained when the robotic arm modified with the fabricated sensor grabbed the "cup". **Fig. 4(b)** shows the classification result when "cup" is the target object. As shown in **Fig. 4(c)**, we analyzed the accuracy using a confusion matrix for nine object recognitions using sensor data. The accuracy was 81.3 % according to the confusion matrix, which is considerably successful. Diagonal elements confirmed that the use of CNT-PDMS nanocomposites for object recognition was possible. Moreover, we investigated the tendency of the accuracy with respect to the number of objects, as shown in **Fig. 4(d)**. It was found that an accuracy > 80 % was obtained for each number of objects using the CNT-PDMS sensor. However, as the number of objects increases, the accuracy decreases because of the loss of the ability of the device to distinguish between different shapes and sizes of objects. Nonetheless, the desired objects were successfully classified using the CNT-PDMS sensor as in an in-sensor computing device. In addition, we analyzed the data obtained to investigate the states that would be effective for in-sensor computing. We divided the region where a transient response was observed into two parts: grabbing and releasing states (shaded red part) at the edge with a static state (blue part) in **Fig. S7(a)**. For this analysis, the number of data points was set to 10. **Fig. S7(b)** shows that when the number of objects increases, the accuracy with dynamic-state data is slightly higher than that with static-state data. Theoretically, smaller datasets are preferable for efficient learning. For instance, in conventional deep-learning techniques, the acquired data are stored in memory and computed using algorithms. However, in reservoir computing devices, there is fading memory that allows the network to conduct learning and prediction tasks without storing big data, which is more efficient. According to **Fig. S7**, using dynamic data, the device can classify nine different objects when 100-step data points are used (**Fig. 5(d)**). In brief, we can classify objects using only dynamic data. This is evident from the reported results; when a robotic hand implemented with our sponge-like sensor grasps an object, it can be classified immediately using a few data points. This would enhance learning efficiency while reducing the power consumption of future reservoir devices.

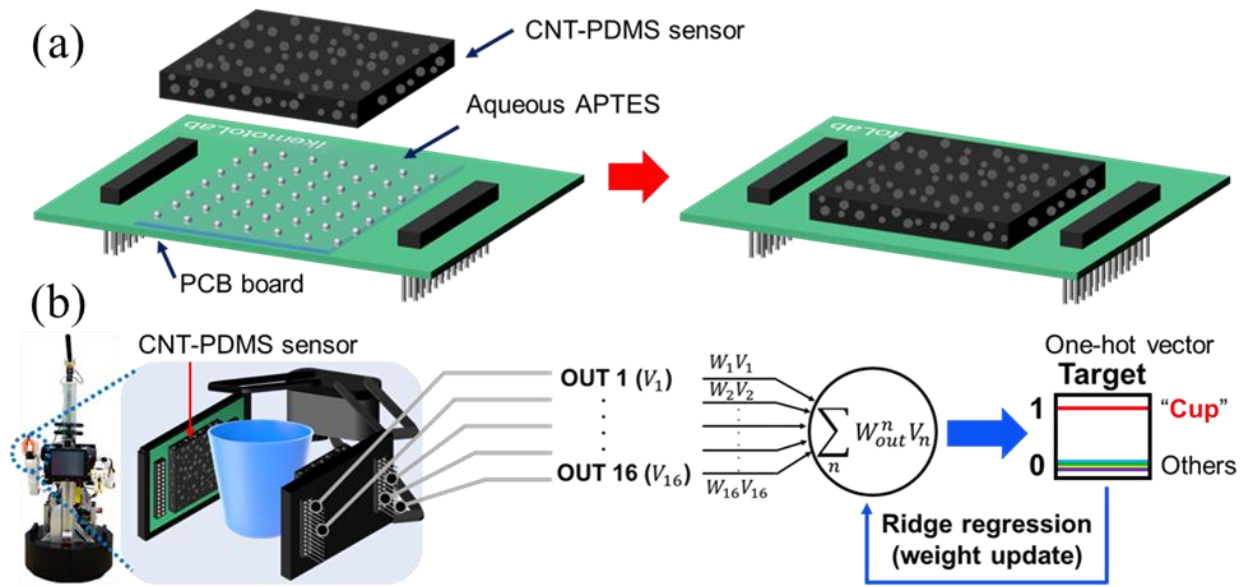

**Fig. 3** (a) Illustration of the attachment method (aqueous (3-Aminopropyl) triethoxysilane (APTES)) for the CNT-PDMS sensor with the printed circuit board. (b) The schematic illustration of object classification using dynamic data.

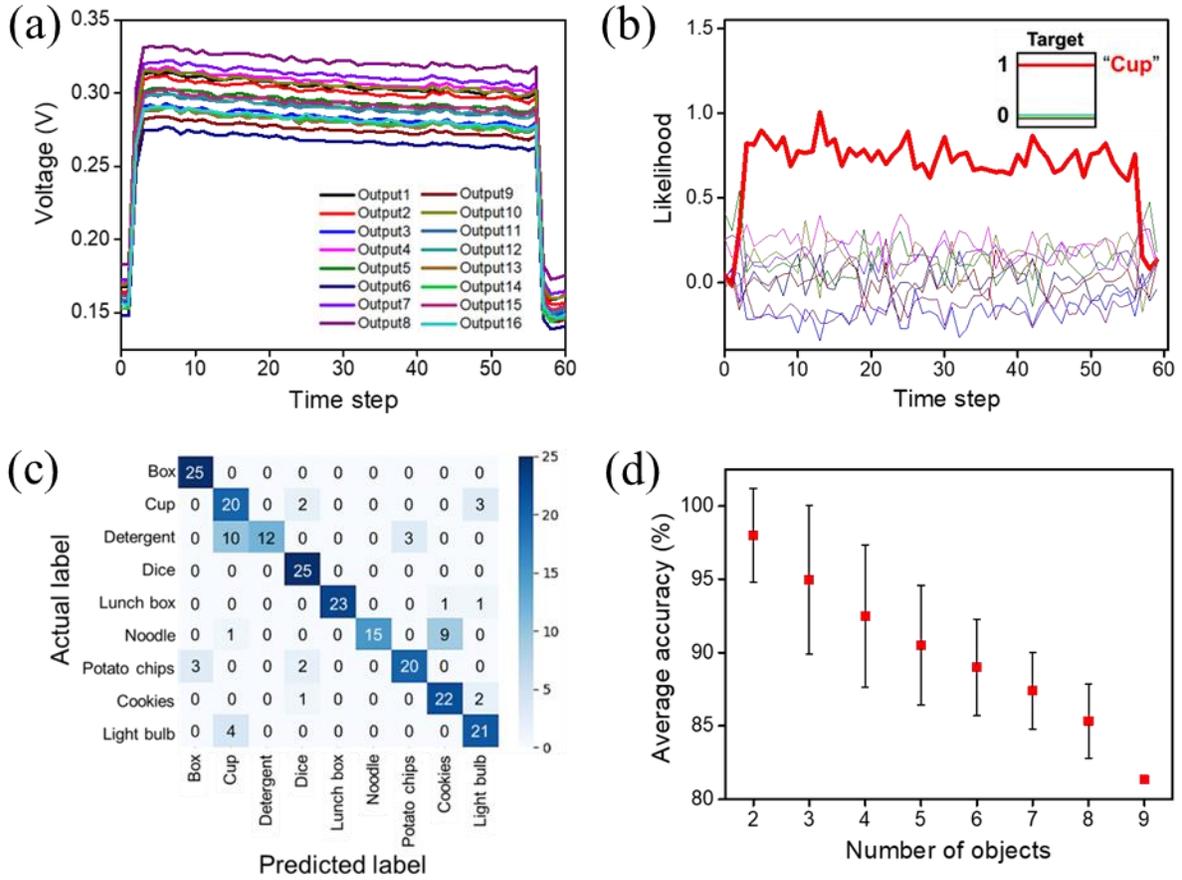

**Fig. 4** (a) Raw data when the cup is grabbed by the CNT-PDMS sensor using robotic hands. (b) Classification results with a cup as a target object. (c) Confusion matrix with nine objects. (d) The tendency of average accuracy by changing the number of objects from two to nine. Error bars represent values obtained by testing all possible combinations.

## Discussion

In this study, we successfully classified nine different objects using a porous piezoresistive sensor composed of CNT/PDMS as an in-sensor computing device using the RC model. Utilizing the sacrificial template method with sugar allows the fabrication of a soft and sensitive sensor. During fabrication, a silane coupling agent helped bond the CNT/PDMS nanocomposite to the polyimide film and PCB board, which improved the adhesion and reduced the contact resistance. Moreover, the change in resistance in response to pressure suggests that the sample with the lowest CNT content has the highest sensitivity. This indicates that a piezoresistive pressure sensor with high sensitivity in the low-pressure range was successfully fabricated, suggesting that fine-tuning certain parameters is crucial for achieving the desired sensor characteristics for specific applications. The recovery time was less than 0.2 s for all the samples, whereas the response time was shorter with a lower amount of CNTs, indicating that the softer sensors responded more quickly. It is important to note that the fabricated sensors consistently generated stable output and exhibited superb stability and endurance throughout our experiments. Consequently, we successfully fabricated a sample with a sufficiently fast recovery time. These results imply that samples with low-CNT content exhibit high sensitivity and fast response times. For the object recognition task, although the classification accuracy decreased as the number of objects increased, accuracy was >80 % for up to nine objects. In conclusion, the fabricated sensor is expected to be used as an "in-sensor reservoir device" to give tactile sensation to robots with artificial intelligence.

## Methods

CNTs have excellent electrical and thermal properties, making them ideal filler materials for enhancing functionality by improving conductivity[26]. PDMS, a type of silicone elastomer, exhibits excellent temperature stability and a low modulus of elasticity that has resulted in its widespread use in contact lenses, microfluidic devices[27,28], and other medical applications[29–31]. By incorporating a small percentage of CNTs into silicone materials, these composite materials can be used as piezoresistive sensors, known as CNT-PDMS sensors[32–36]. The sacrificial-template-based fabrication method has been utilized to fabricate porous structures that are expected to be simple and cost-effective without the need for advanced equipment. PDMS prepolymer (Sylgard 184 Silicone Elastomer, Dow Corning, Midland, MI, USA) was prepared by mixing 30 g of base and 3 g of curing agents; the mixture was vigorously stirred for 20 min, covered to avoid contamination, and kept in the freezer at -20 °C to avoid polymerization while degassing[25]. Various weight fractions ($x$ = 0.2, 0.3, 0.4, and 0.5 g) of Multi-Walled CNTs (MWCNTs, FT9110, Cnano Technology, Santa Clara, CA, USA)) were directly mixed with 20 g of sugar using a mortar and pestle. The resulting mixture was then wetted with 600 µL of distilled water and molded into a 5 × 5 × 0.5 cm³ cuboid. After drying the sample in the oven at 70 °C for 4 h, the MWCNTs-sugar cuboid was placed in a Petri dish filled with previously prepared PDMS prepolymer. The Petri dish was placed in a vacuum desiccator for 1 h at -0.1 MPa below atmospheric pressure. The sample was then covered to avoid contamination and placed in the freezer at -12.4 °C for an additional hour. This process resulted in PDMS infusion into the MWCNT-sugar pores owing to the capillary effect caused by the vacuum and shrinkage. The MWCNTs-sugar-PDMS cuboid was then taken out, excess PDMS prepolymer on the surface was wiped off, and the sample was placed in the oven at 70 °C for 24 h to cure the PDMS. After curing, the sample surfaces were polished with sandpaper. The sample was then placed in a beaker filled with distilled water and placed in a conventional microwave oven for 5 min to remove sugar. Microwave heating expedited sugar dissolution owing to pore expansion[37]. The dissolution of sugar in the water changed the color of the water to brown. Water was discarded after each cycle. This process was repeated five or more times until the water remained clear. After drying, we obtained a CNT-PDMS nanocomposites (**Fig. 5(a)**). Subsequently, we fabricated a sensor device to evaluate the mechanical and electrical behavior of nanocomposite with different CNT contents (0.2 g to 0.5 g). The CNT-PDMS sensor device was fabricated by cutting the MWCNTs-PDMS nanocomposite into 1.5 × 1.5 × 0.5 cm³ pieces using a razor blade and attaching them to screen-printed interdigitated silver electrodes on a polyimide film (Kapton). For attachment, a silane coupling agent (1 vol % aqueous (3-Aminopropyl) triethoxysilane (APTES), Sigma–Aldrich) was employed to reduce the effects of contact resistance[25]. APTES is an aminosilane compound that combines both the reactive functional group ("amino group"), which is expected to react and interact with organic materials, and the hydrolyzable alkoxysilane group[38]. Silane coupling was performed by first placing the electrodes and MWCNT–PDMS in an ultraviolet ozone cleaner for 30 min, and then surface-functionalizing the electrodes by placing them in 1 vol % aqueous APTES for 30 min. Lastly, MWCNTs-PDMS nanocomposite was placed on top of the electrode, clamped tightly, and placed in the oven at 70 °C for 1 h. **Fig. 5(b)** represents the CNT-PDMS sensor device. These sensors were used to investigate the sensitivity, response time, and recovery time of the CNT-PDMS nanocomposites when an external force was applied. The sensitivity of the sensor was characterized using the setup shown in **Fig. 5(c)**. The sensors were attached to a standard digital force gauge (ZTS-5N, IMADA Inc, Norhbrook, Illinois, USA) using double-sided tape, and the force was applied using a micrometer (Mitsutoyo digital micrometer: MHN3-25MB, Mitsutoyo, Kawasaki, Kanagawa, Japan). A three-dimensional-printed plate was placed between the micrometer and the sample to cover the entire sample area. The micrometer was manually turned by 0.05 mm at each rotational step, while the output voltage was recorded using an Arduino MEGA2560 and was converted to resistance after measurement. The input voltage of the sensor was set to 5 V using an Arduino kit. The displacement and voltage values were recorded using MATLAB. Furthermore, to confirm the stability and durability of the fabricated sensor, we conducted cyclic measurements with a force gauge stand machine (MX2-500N Motorized Test Stand, IMADA Inc, Northbrook, Illinois, USA) by applying a pressure 10000 times (the start and return speeds were configured at 200 mm/min, while the measuring speed was set to 100 mm/min). Subsequently, we recorded the output voltage to observe the differences between the different states throughout the measurement using MATLAB (R2024a, MathWorks, Natick, Massachusetts, USA).

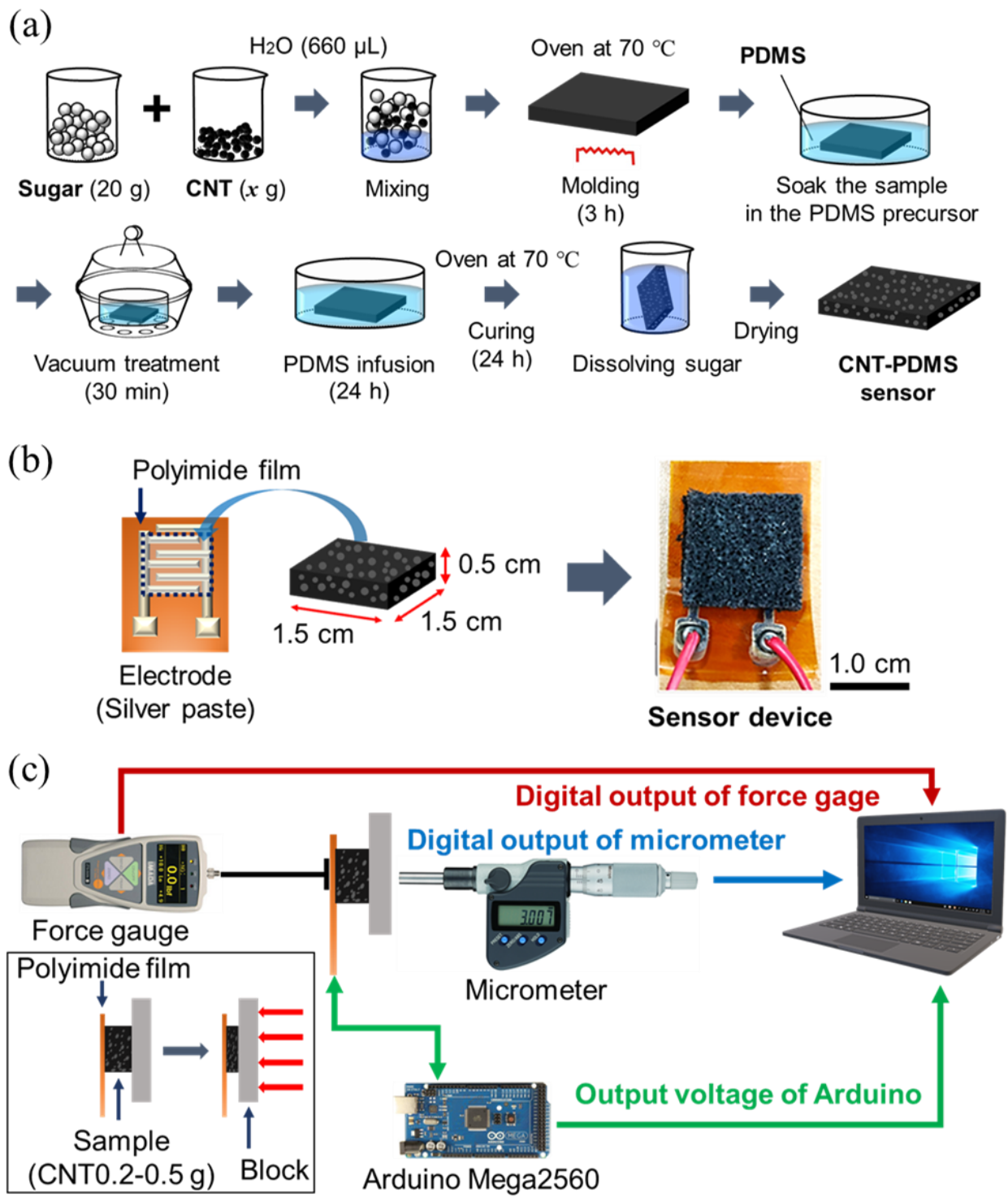

**Fig. 5** (a) Schematic illustration of the carbon nanotube-dimethylpolysiloxane (CNT-PDMS) nanocomposite preparation. (b) Schematic and image of the fabricated device. (c) Measurement set up for evaluating sensitivity *S*. Force gauge, micrometer, and Arduino Mega2560 were used to record the respective outputs of force (N), displacement (mm), and voltage (V).

References

1. Wan, T. *et al.* In-sensor computing: Materials, devices, and integration technologies. *Adv. Mater.* 35(37), e2203830 (2023). doi:10.1002/adma.202203830.
2. Sun, L. *et al.* In-sensor reservoir computing for language learning via two-dimensional memristors. *Sci. Adv.* 7, 1455–1469 (2021).
3. Liu, L., Zhang, Y. & Yan, Y. Four levels of in-sensor computing in bionic olfaction: from discrete components to multi-modal integrations. *Nanoscale Horiz.* 8, 1301–1312 (2023). https://doi.org/10.1039/d3nh00115f.
4. Jiang, C. *et al.* 60 nm Pixel-size pressure piezo-memory system as ultrahigh-resolution neuromorphic tactile sensor for in-chip computing. *Nano Energy* 87(5), 106190 (2021). https://doi.org/10.1016/j.nanoen.2021.106190.
5. Zhang, H. *et al.* In-sensor computing realization using fully CMOS-compatible TiN/HfOx-based neuristor array. *ACS Sens.* 8, 3873–3881 (2023).
6. Tanaka, G. *et al.* Recent advances in physical reservoir computing: A review. *Neural Netw.* 115, 100–123 (2019).
7. Tanaka, H. *et al.* In-materio computing in random networks of carbon nanotubes complexed with chemically dynamic molecules: a review. *Neuromorphic Comput. Eng.* 2, 022002 (2022).
8. Banerjee, D. *et al.* Emergence of in-materio intelligence from an incidental structure of a single-walled carbon nanotube–porphyrin polyoxometalate random network. *Adv. Intell. Syst.* 4, 2100145 (2022).
9. Kotooka, T., Tanaka, Y., Tamukoh, H., Usami, Y. & Tanaka, H. Random network device fabricated using Ag2Se nanowires for data augmentation with binarized convolutional neural network. *Appl. Phys. Express* 16, 014002 (2023).
10. Usami, Y. *et al.* In-materio reservoir computing in a sulfonated polyaniline network. *Adv. Mater.* 33(48), e2102688 (2021).
11. Stassi, S., Cauda, V., Canavese, G. & Pirri, C. F. Flexible tactile sensing based on

piezoresistive composites: A review. *Sensors* 14, 5296–5332 (2014).
12. Boutry, C. M. *et al.* A stretchable and biodegradable strain and pressure sensor for orthopaedic application. *Nat. Electron.* 1, 314–321 (2018).
13. Huang, H. *et al.* Tunable thermal-response shape memory bio-polymer hydrogelsas body motion sensors. *Eng. Sci.* 9, 60–67 (2020).
14. Qu, M. *et al.* Flexible conductive Ag-CNTs sponge with corrosion resistance for wet condition sensing and human motion detection. *Colloid. Surf. A Physicochem. Eng. Asp.* 656, 130427 (2023).
15. Yogeswaran, N. *et al.* Tuning electrical conductivity of CNT-PDMS nanocomposites for flexible electronic applications. *IEEE-NANO 2015 - 15th Int. Conf. Nanotechnol.* 1441–1444 (2015) doi:10.1109/NANO.2015.7388911.
16. Park, S., Vosguerichian, M. & Bao, Z. A review of fabrication and applications of carbon nanotube film-based flexible electronics. *Nanoscale* 5, 1727–1752 (2013).
17. Gong, S. *et al.* A wearable and highly sensitive pressure sensor with ultrathin gold nanowires. *Nat. Commun.* 5, 1–8 (2014).
18. Han, M. *et al.* Highly sensitive and flexible wearable pressure sensor with dielectric elastomer and carbon nanotube electrodes. *Sens. Actuators A Phys.* 305, 111941 (2020).
19. Song, Y. *et al.* Highly compressible integrated supercapacitor–piezoresistance-sensor system with CNT–PDMS sponge for health monitoring. *Small* 13, 1–10 (2017).
20. Lim, C. *et al.* Stretchable conductive nanocomposite based on alginate hydrogel and silver nanowires for wearable electronics. *APL Mater.* 7, 031502 (2018).
21. Su, Y. F., Han, G., Kong, Z., Nantung, T. & Lu, N. Embeddable piezoelectric sensors for strength gain monitoring of cementitious materials: The influence of coating materials. *Eng. Sci.* 11, 66–75 (2020).
22. Azhari, S. *et al.* Toward automated tomato harvesting system: Integration of haptic based piezoresistive nanocomposite and machine learning. *IEEE Sens. J.* 21, 27810–27817 (2021).
23. Iijima, S. Helical microtubules of graphitic carbon. *Nature* 354, 56–58 (1991).
24. Fujii, T. PDMS-based microfluidic devices for biomedical applications. *Microelectron. Eng.* 61–62, 907–914 (2002).
25. Martínez-Brenes, A. *et al.* Combined electrokinetic manipulations of pathogenic bacterial

samples in low-cost fabricated dielectrophoretic devices. *AIP Adv.* 9, (2019).

26. Lin, C. H., Yeh, Y. H., Lin, W. C. & Yang, M. C. Novel silicone hydrogel based on PDMS and PEGMA for contact lens application. *Colloid. Surf. B Biointerfaces* 123, 986–994 (2014).
27. Tran, N. P. D. & Yang, M. C. The ophthalmic performance of hydrogel contact lenses loaded with silicone nanoparticles. *Polymers (Basel).* 12, 1128 (2020).
28. Chen, J. S. *et al.* Biopolymer brushes grown on PDMS contact lenses by in situ atmospheric plasma-induced polymerization. *J. Polym. Res.* 24, 1–9 (2017).
29. Azhari, S. *et al.* Fabrication of piezoresistive based pressure sensor via purified and functionalized CNTs/PDMS nanocomposite: Toward development of haptic sensors. *Sens. Actuators A Phys.* 266, 158–165 (2017).
30. Kim, J. H. *et al.* Simple and cost-effective method of highly conductive and elastic carbon nanotube/polydimethylsiloxane composite for wearable electronics. *Sci. Rep.* 81(8), 1–11 (2018).
31. Ahir, S. V., Huang, Y. Y. & Terentjev, E. M. Polymers with aligned carbon nanotubes: Active composite materials. *Polymer (Guildf).* 49, 3841–3854 (2008).
32. Oser, P. *et al.* Fiber-optic photoacoustic generator realized by inkjet-printing of CNT-PDMS composites on fiber end faces. *Macromolecular Mater. Eng.* 306(2), 2000563 (2000). doi:10.1002/mame.202000563.
33. Zhang, Y. *et al.* Polymer-embedded carbon nanotube ribbons for stretchable conductors. *Adv. Mater.* 22, 3027–3031 (2010).
34. Azhari, S. *et al.* Integration of wireless power transfer technology with hierarchical multiwalled carbon nanotubes-polydimethylsiloxane piezo-responsive pressure sensor for remote force measurement. *IEEE Sens. J.* 23, 7902–7909 (2023).
35. Herren, B. *et al.* PDMS sponges with embedded carbon nanotubes as piezoresistive sensors for human motion detection. *Nanomaterials* 11(7), 1740 (2021).
36. Acres, R. G. *et al.* Molecular structure of 3-aminopropyltriethoxysilane layers formed on silanol-terminated silicon surfaces. *J. Phys. Chem. C* 116, 6289–6297 (2012).
37. Yu, Q. *et al.* Porous pure MXene fibrous network for highly sensitive pressure sensors. *Langmuir* 38, 5494–5501 (2022).
38. Jung, Y., Jung, K. K., Kim, D. H., Kwak, D. H. & Ko, J. S. Linearly sensitive and flexible

pressure sensor based on porous carbon nanotube/polydimethylsiloxane composite structure. *Polymers* 12(7), 1499. doi:10.3390/polym12071499.

39. Zhou, F. & Chai, Y. Near-sensor and in-sensor computing. *Nat. Electron.* 3, 664–671 (2020).

40. Tanaka, H. *et al.* Sensors for Mechanical Quantities and Their Integration into Robotics: Sensing Approaches and Sensor Fabrication Processes. Patent No.21064, 2022-071679 (2022).

41. Tanaka, H. *et al.* Recognition System and Robot Integration. Patent No.2023-115390 (2023).

## Acknowledgements

This study was technologically supported by the Kitakyushu Semiconductor Center under the "Advanced Research Infrastructure for Materials and Nanotechnology in Japan (ARIM Japan)" of the Ministry of Education, Culture, Sports, Science, and Technology (MEXT), Japan. This study was financially supported by the Japanese Society for the Promotion of Science (JSPS) KAKENHI [grant numbers 19H02559, 19K22114, 20K21819, 21K14527, 22H01900, 23K17864, and 23K18495] and Japan Science and Technology Agency (JST) CREST [grant number JPMJCR21B5] and ACT-X [grant number JPMJAX22K4], and ALCA-Next [grant number JPMJAN23F3], and JSPS Core-to-Core Project [grant number JPJSCCA20220006]. S.A. and Y.U. thank Asahi Kohsan Co., Ltd. for the financial support provided by the Kitakyushu Foundation for the Advancement of Industry, Science, and Technology, Japan. K.K. thanks JST for the university fellowships awarded to promote the creation of science and technology innovation [grant number JPMJFS2133].

Supplementary Information

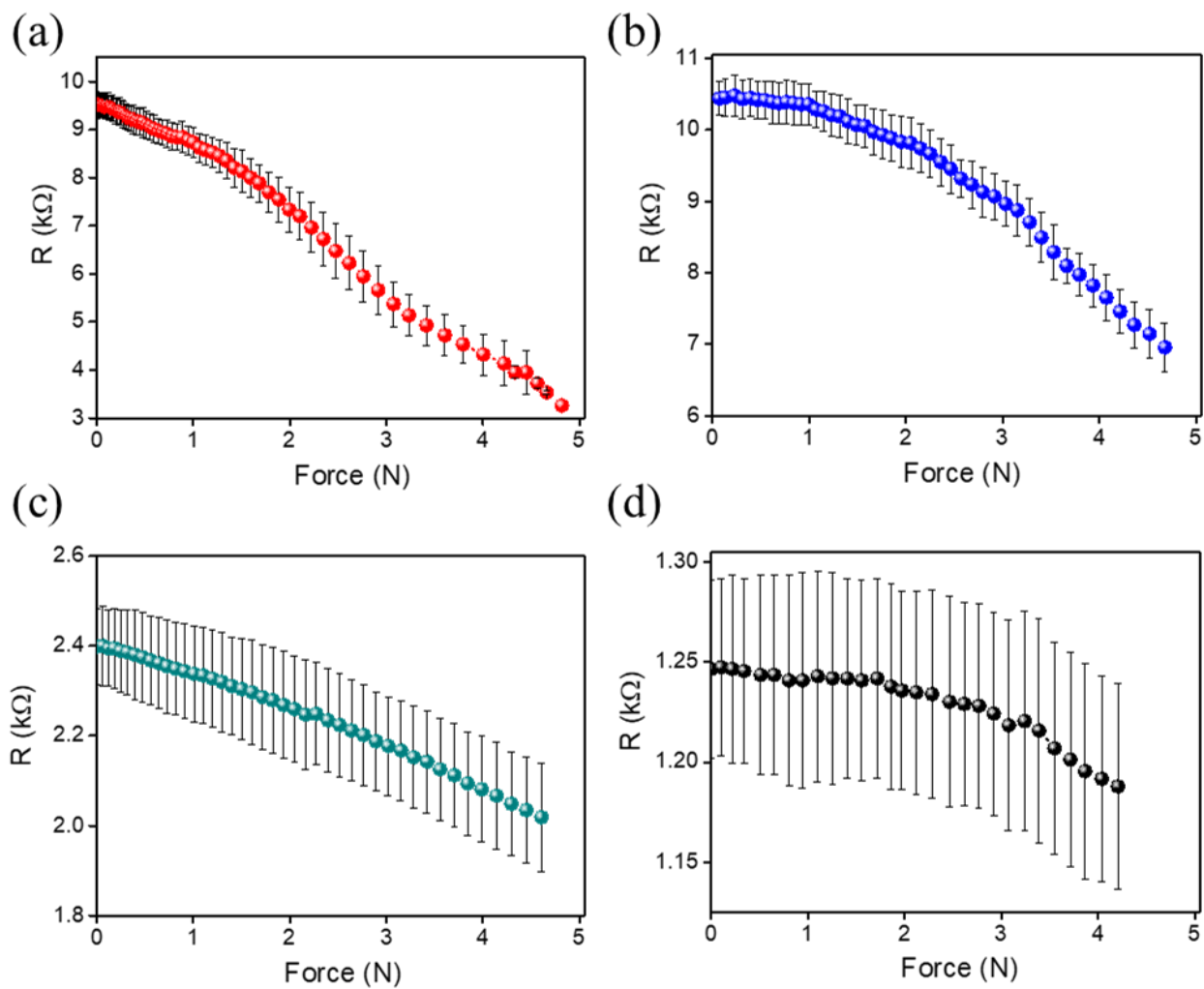

**Fig. S1** Graphs of force and resistance when force was applied to carbon nanotube-dimethylpolysiloxane (CNT-PDMS) sensors. (a)-(d) CNT: 0.2–0.5 g.

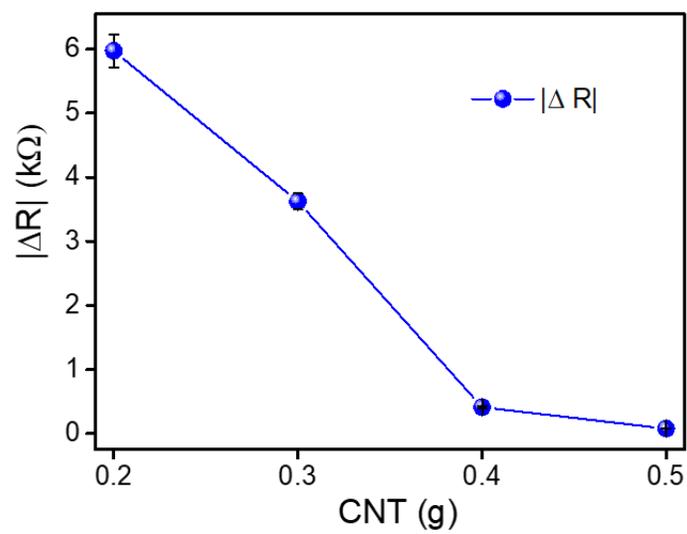

**Fig. S2** Relative change in sensor's resistance as a function of CNT content.

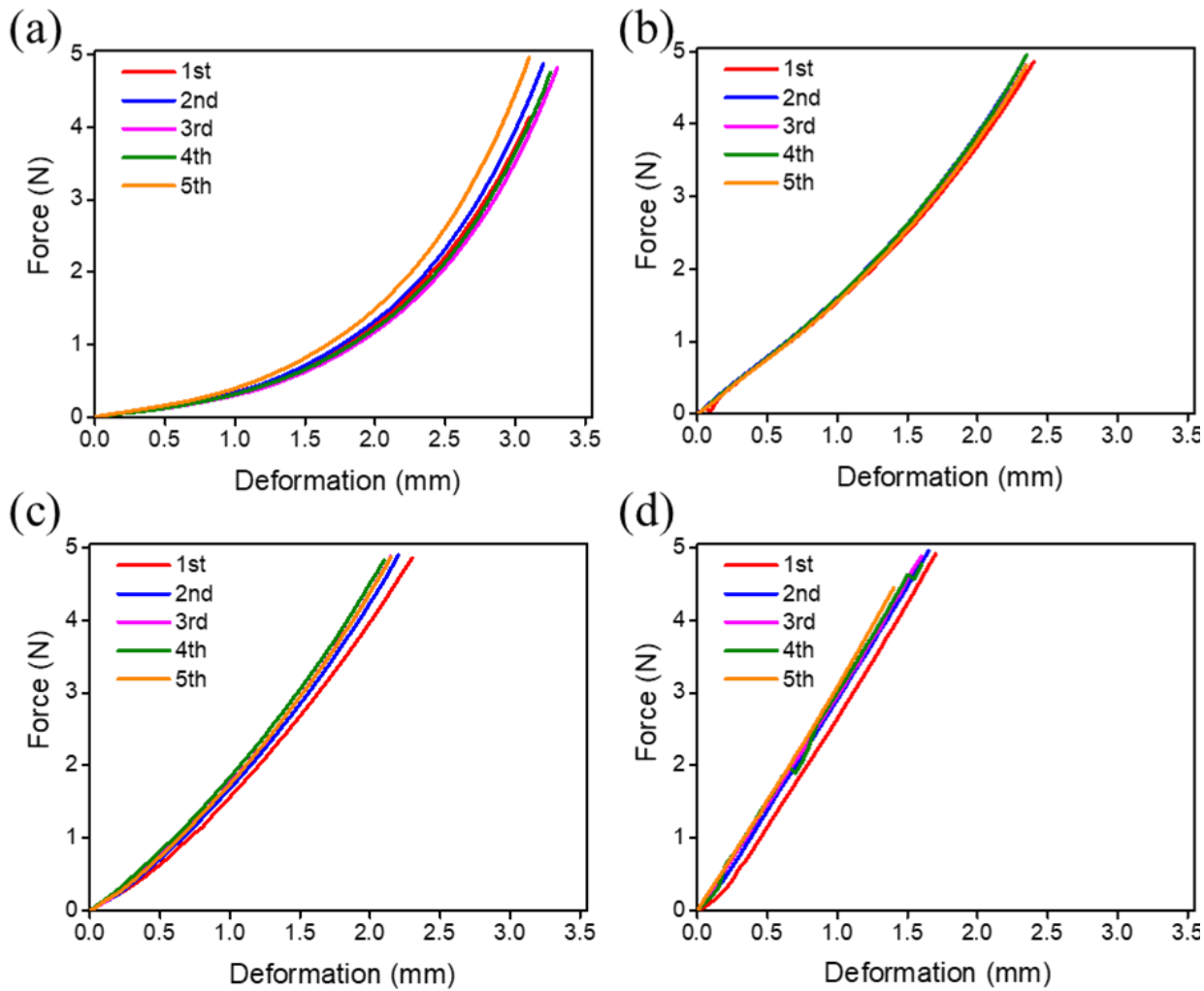

**Fig. S3** Force change curves of CNT-PDMS nanocomposites as a function of deformation when force is applied. (a)-(d) CNT: 0.2–0.5 g.

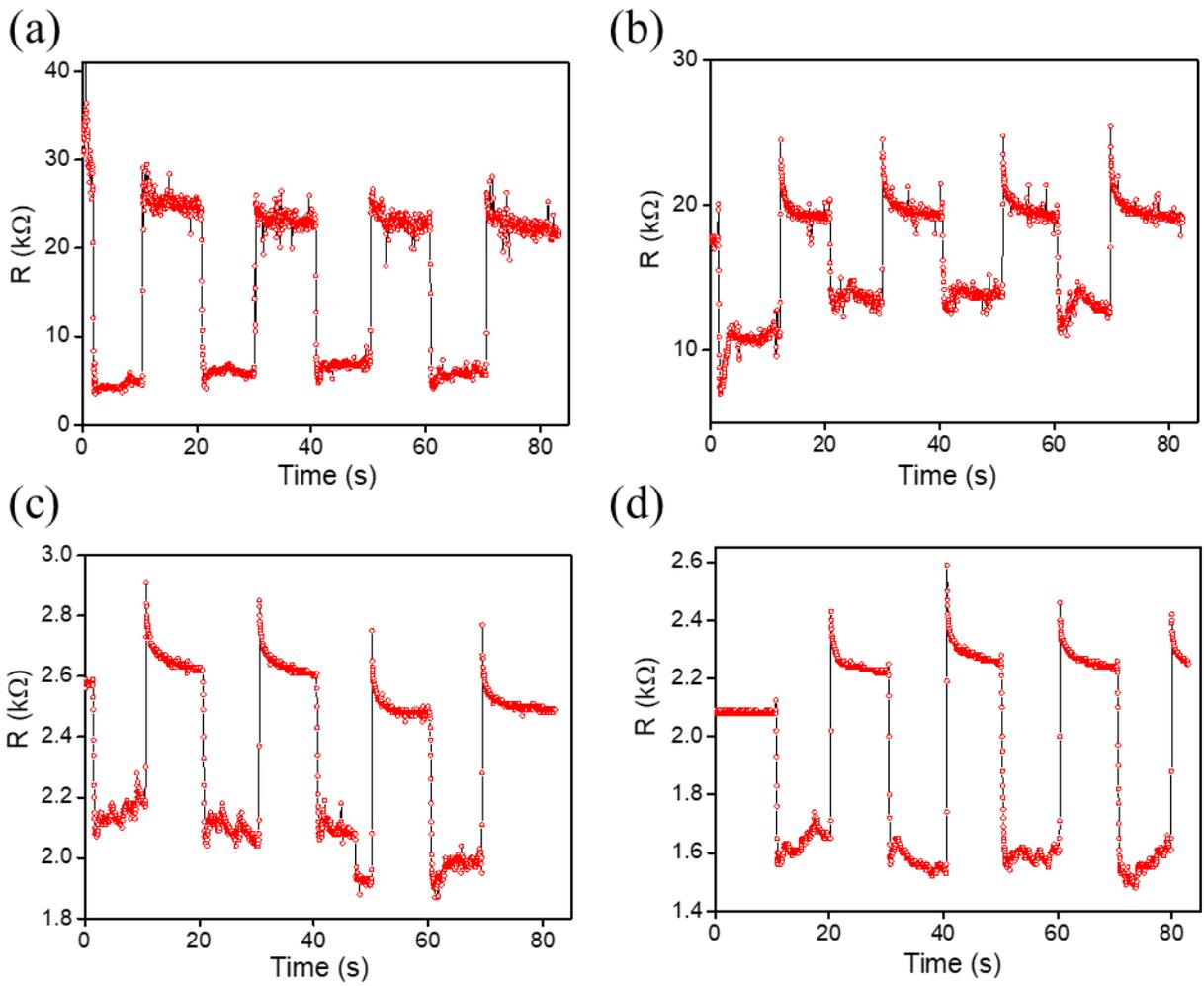

**Fig. S4** (a)-(d) Repeated measurements of pressure application and unloading (CNT: 0.2–0.5 g).

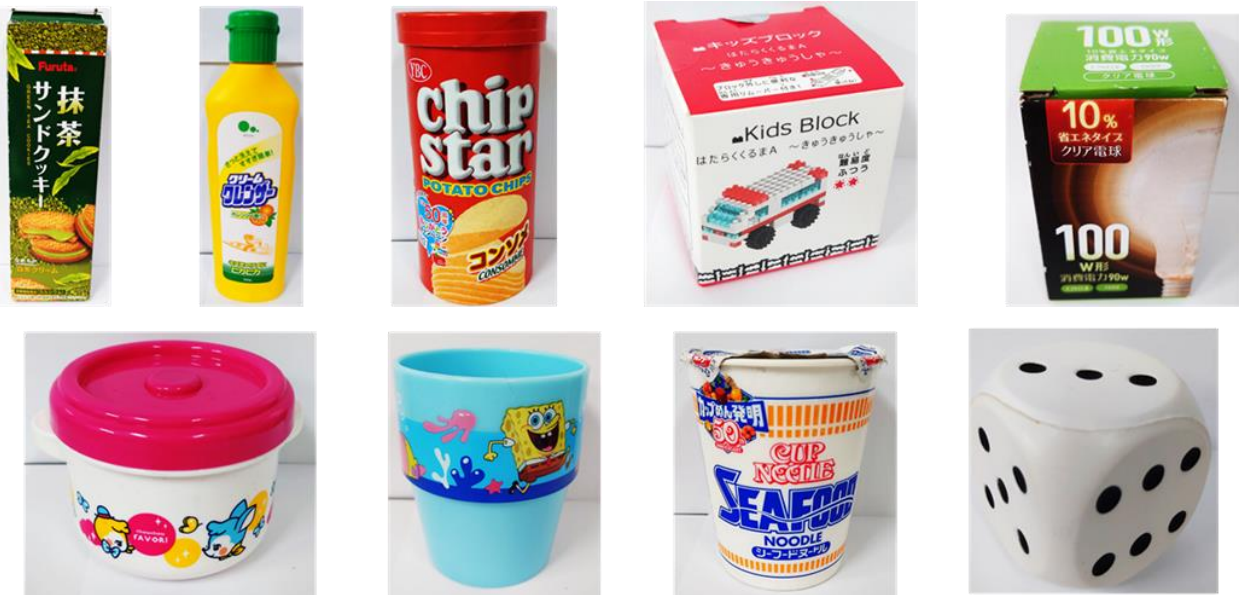

**Fig. S5** Nine objects used for classification (cookie box, detergent, snack can, toy box, light bulb box, lunch box, cup, noodle, dice).

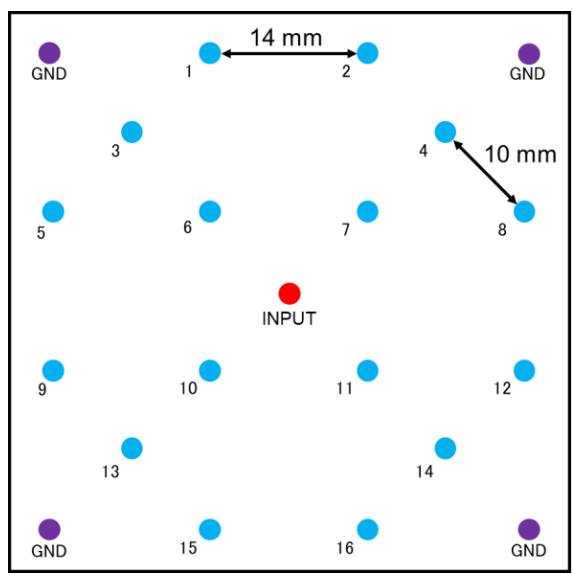

**Fig. S6** Schematic of PCB board. (The input of 5 V is positioned in the center, with 16 outputs distributed evenly around it, and the ground (GND) is located at the corner)

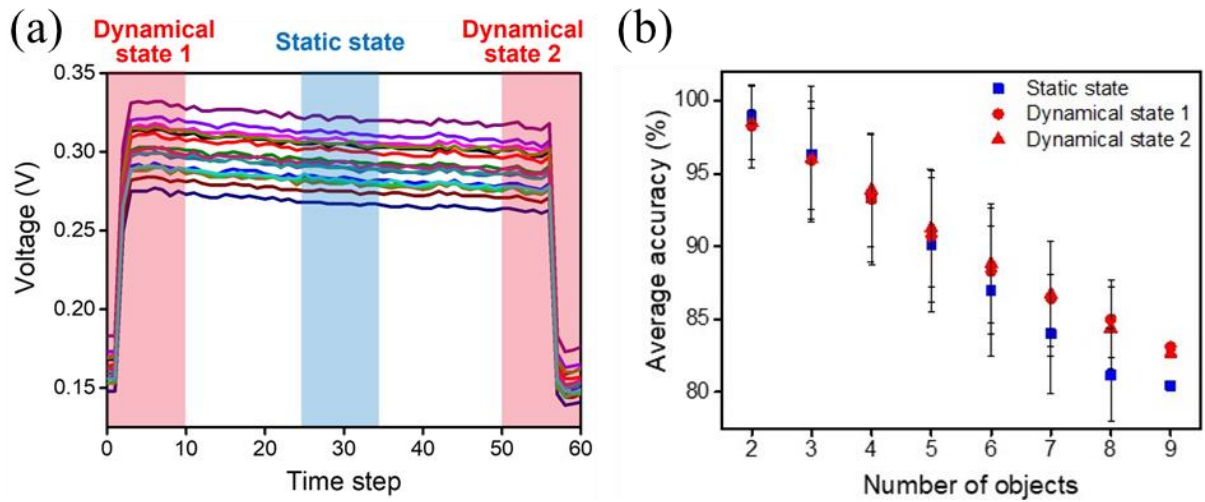

**Fig. S7** (a) Obtained data were divided into regions with dynamical states 1 and 2 (shaded red part) and a region showing the static state (blue part). (b) The tendency of average accuracy by changing the number of objects from 2 to 9. Error bars represent values obtained by testing all possible combinations.